# Tracking fast and slow changes in synaptic weights from simultaneously observed pre- and postsynaptic spiking

Ganchao Wei[1*] and Ian H. Stevenson[2,3]

[1]Department of Statistics, University of Connecticut
[2]Department of Psychological Sciences, University of Connecticut
[3]Department of Biomedical Engineering, University of Connecticut

[*]Corresponding Author: ganchao.wei@uconn.edu

## Abstract

Synapses change on multiple timescales, ranging from milliseconds to minutes, due to a combination of both short- and long-term plasticity. Here we develop an extension of the common Generalized Linear Model to infer both short- and long-term changes in the coupling between a pre- and post-synaptic neuron based on observed spiking activity. We model short-term synaptic plasticity using additive effects that depend on the presynaptic spike timing, and we model long-term changes in both synaptic weight and baseline firing rate using point process adaptive smoothing. Using simulations, we first show that this model can accurately recover time-varying synaptic weights 1) for both depressing and facilitating synapses, 2) with a variety of long-term changes (including realistic changes, such as due to STDP), 3) with a range of pre- and post-synaptic firing rates, and 4) for both excitatory and inhibitory synapses. We then apply our model to two experimentally recorded putative synaptic connections. We find that simultaneously tracking fast changes in synaptic weights, slow changes in synaptic weights, and unexplained variations in baseline firing is essential. Omitting any one of these factors can lead to spurious inferences for the others. Altogether, this model provides a flexible framework for tracking short- and long-term variation in spike transmission.

# 1 Introduction

Detecting and characterizing synaptic connections and how they change over time is a major experimental challenge, especially in behaving animals. Although many studies of synaptic transmission focus on intracellular recordings, in some cases, monosynaptic connections can be identified in extracellular spike recordings (Barthó et al., 2004; Fetz et al., 1991). When there is a synaptic connection, the probability of the postsynaptic neuron spiking will briefly increase or decrease after presynaptic spikes, for an excitatory or inhibitory synapse, respectively. Here, a synaptic connection can sometimes be detected in the empirical cross-correlation or cross-correlogram between pre- and postsynaptic spiking (Perkel et al., 1967). A transient, short latency peak or trough in the cross-correlogram can suggest the presence of a synapse. However, the accuracy with which synapses can be detected is limited, especially with short recordings from weakly connected, sparsely firing neurons. Recently, model-based methods have been shown to improve detection accuracy (Kobayashi et al., 2019; Ren et al., 2020).

In cases where synapses can be reliably identified from spikes, these recordings can potentially be used to examine changes in synaptic strength over time (Fujisawa et al., 2008; McKenzie et al., 2021). Changes in synaptic strength occur over multiple timescales and due to different biophysical mechanisms (Zucker & Regehr, 2002). For instance, on timescales of a few milliseconds short-term synaptic plasticity (STSP) occurs, where the synaptic strength may decrease due to vesicle depletion (depression) or increase due to the influx of calcium (facilitation). In contrast, on timescales of minutes to hours long-term synaptic plasticity (LTSP) occurs, where the synaptic strength increases (long-term potentiation, LTP) or decreases (long-term depression, LTD) due to changes in receptor density or structure. It is important to note that the effects of STSP can appear on longer timescales due to fluctuations in the presynaptic rate (Kandaswamy et al., 2010; Klyachko & Stevens, 2006). For instance, in synapses with short-term depression, high presynaptic rates lead to a chronically depleted state, and the synapses may appear stronger when the rate decreases. Short- and long-term changes in synaptic strength can be roughly studied by examining the cross-correlations corresponding to different presynaptic patterns, such as inter-spike intervals (ISI) (Carandini et al., 2007; Csicsvari et al., 1998; Fujisawa et al., 2008; Mantel & Lemon, 1987; Swadlow & Gusev, 2001; Usrey et al., 2000) or different recording periods. However, cross-correlograms for any given partition will be affected by both short- and long-term effects and how to best split the cross-correlograms is often unclear.

Here we model both timescales simultaneously and decompose the synaptic weight into a short- and long-term effect using a model-based approach (an extension of a Generalized Linear Model, GLM). The STSP term allows depression/facilitation to be modeled as a function of the recent presynaptic inter-spike intervals (ISIs) (Ghanbari et al., 2017), and we track nonspecific changes in LTSP using point-process adaptive smoothing (Eden et al., 2004). We also model fluctuations in the baseline postsynaptic rate that could potentially influence the inference of synaptic strength. Several previous studies have proposed models for estimating either long-term (Linderman et al., 2014; Song et al., 2018; Stevenson & Kording, 2011) or short-term (Chan et al., 2008; English et al., 2017; Ghanbari et al., 2017) changes in synaptic weights from pre- and postsynaptic spiking. Our work builds on these approaches and provides a flexible framework for tracking both short- and long-term variation in spike transmission. Here we present results from several simulations, as well as two experimentally recorded putative synaptic connections, and demonstrate why estimating both long-term and short-term effects simultaneously is necessary for accurate inference.

## 2 Methods

Here we introduce an extension of a Poisson GLM that aims to describe the coupling between a pre- and postsynaptic neuron (Brillinger, 1988, 1992). While many previous studies have modeled static coupling between neurons (Harris et al., 2003; Okatan et al., 2005; Pillow et al., 2008; Truccolo et al., 2005), our goal here is to describe a time-varying synaptic strength with both short- and long-term changes.

We model the postsynaptic spiking in discrete time as a doubly stochastic Poisson process with time-varying parameters. Partitioning the total recording time T into evenly-spaced bins $\{t_k\}_{k=0}^{N}$, such that $0 = t_0 \leq \cdots \leq t_N = T$ with time steps $\Delta t$, we denote the total number of presynaptic spikes in $t \in (0, t_k]$ as $N_k^{pre}$, and $y_k^{pre} = N_k^{pre} - N_{k-1}^{pre}$ represents the number of spikes observed in $t \in (t_{k-1}, t_k]$. Similarly, $N_k^{post}$ denotes the total number of postsynaptic spikes in $t \in (0, t_k]$, and $y_k^{post}$ denotes the number of postsynaptic spikes observed in $t \in (t_{k-1}, t_k]$. For small enough $\Delta t$ both $y_k^{pre}$ and $y_k^{post}$ take values of 0 or 1 and can be viewed as spike indicators for time bin $k$.

Previous models of static coupling between neurons typically include the recent spiking history of both the presynaptic input (the coupling effect) and the postsynaptic neuron itself.

$$\lambda_k = \lambda\big(t_k \,\big|\, \boldsymbol{y}_{1:k-1}^{post}, \boldsymbol{y}_{1:k-1}^{pre}\big) = \exp\big(\beta_0 + \boldsymbol{h}\boldsymbol{y}_{k-h:k-1}^{post} + \boldsymbol{g}\boldsymbol{y}_{k-h:k-1}^{pre}\big) \quad (2.1)$$

$$y_k^{post} \sim Poisson(\lambda_k \Delta t) \quad (2.2)$$

where $\lambda_k$ is the conditional intensity of the postsynaptic neuron at $t_k$, given the recent spiking of the postsynaptic neuron $\boldsymbol{y}_{k-h:k-1}^{post}$ and presynaptic neuron $\boldsymbol{y}_{k-h:k-1}^{pre}$, $h$ steps into the past. For the model parameters, $\beta_0$ defines the fixed baseline firing rate and the coupling and history effects are weighted by $\boldsymbol{h}$ and $\boldsymbol{g}$, for the post- and pre-synaptic neurons, respectively.

Here we extend this static model to include fast and slow dynamics in the coupling effect and model the postsynaptic neuron's firing rate as

$$\lambda_k = \lambda\big(t_k \big| \boldsymbol{y}_{1:k-1}^{post}, \boldsymbol{y}_{1:k-1}^{pre}\big) = \exp\big(\beta_0(t_k) + \boldsymbol{h}\boldsymbol{y}_{k-h:k-1}^{post} + w(t_k) \cdot x(t_k)\big)$$
$$= \exp\big(\beta_0(t_k) + \boldsymbol{h}\boldsymbol{y}_{k-h:k-1}^{post} + w_L(t_k)w_S(t_k) \cdot x(t_k)\big) \quad (2.3)$$

$$y_k^{post} \sim Poisson(\lambda_k \Delta t) \quad (2.4)$$

In our model, the conditional intensity of the postsynaptic neuron is captured by an unstructured time-varying baseline $\beta_0(t_k)$, a history effect $\boldsymbol{h}\boldsymbol{y}^{post}_{k-h:k-1}$, and a time-varying coupling term $w(t_k) \cdot x(t_k)$. In the static coupling model, the main goal is to accurately infer the shapes of the filters, here we assume that the shape is fixed $x(t_k) = \boldsymbol{g}\boldsymbol{y}^{pre}_{1:k-1}$, but is weighted by an additional factor $w(t_k)$ that varies over time. Additionally, we aim to partition the variations in $w_k$ into a long-term component $w_L(t_k)$ and a short-term component $w_S(t_k)$ that together determine the synaptic weight. To account for short-term synaptic facilitation/depression, we model a transient increase/decrease in $w_S(t_k)$ after each presynaptic spike, where the amplitude of the increase/decrease depends on the preceding presynaptic inter-spike interval. In the absence of presynaptic activity, $w_S(t_k)$ returns to a base value of 1. To simplify the notation, we write $\beta_0(t_k)$, $w_L(t_k)$, $w_S(t_k)$ and $x(t_k)$ as $\beta_{0,k}$, $w_{L,k}$, $w_{S,k}$ and $x_k$ below.

Estimating the Synaptic Filter

Here we estimate the shape of the static presynaptic filter $g$ by directly modeling the cross-correlogram between the pre- and postsynaptic spikes, similar to Ren et al. (Ren et al., 2020). Briefly, we assume that the shape of the synaptic connection is described by an alpha function $g(\cdot) = \alpha(t, \Delta t_\alpha, \tau_\alpha) = \frac{t - \Delta t_\alpha}{\tau_\alpha} \exp\left(1 - \frac{t - \Delta t_\alpha}{\tau_\alpha}\right) I(t > \Delta t_\alpha)$, with a latency $\Delta t_\alpha$ and time constant $\tau_\alpha$. These parameters are then estimated by modeling the cross-correlogram $z_c$ as a combination of a slow background correlation and a fast, transient effect of the synaptic connection:

$$\lambda_{cross,m} = \exp\left(\alpha_0 + X_{c,m}\alpha_c + w_{cross}\,\alpha(m, \Delta t_\alpha, \tau_\alpha) * a_m\right) \quad (2.5)$$
$$z_{c,m} \sim Poisson(\lambda_{cross,m}) \quad (2.6)$$
$$z_{c,m} = \sum_t y^{pre}_{t-m} \cdot y^{post}_t \text{ and } a_m = \sum_t y^{pre}_{t-m} \cdot y^{pre}_t \quad (2.7)$$

where $\lambda_{cross,m}$ denotes the expected rate of coincidences from the cross-correlogram with baseline $\alpha_0$ and a linear combination of $c$ smooth basis functions $X_{c,m}\alpha_c$ accounting for slow changes in cross-correlation, at time bin $m$. Here we use $c = 4$ cubic B-spline bases with equally spaced knots for the smooth basis $X$ and model the time-range $[-50\text{ms}, 50\text{ms}]$ in $1\text{ms}$ bins. The fast effect is described as $w_{cross}\ \alpha(m, \Delta t_\alpha, \tau_\alpha)$ where $w_{cross}$ is the connection strength from pre-synaptic neuron to post-synaptic neuron, and the alpha function is convolved with the autocorrelation of the presynaptic neuron $a_m$ to account for the effects of presynaptic dynamics. The parameters are estimated by maximizing the Poisson log-likelihood. We use random restarts, since the objective function is non-convex in the latency and time-constant parameters.

After fitting the latency and time constant of the synaptic filter to the cross-correlogram we assume that these parameters are fixed when modeling the long- and short-term changes in synaptic strength. This simplifying assumption allows us to model rescaling of the basic presynaptic input, given by $x_k = \alpha(t, \Delta t_\alpha, \tau_\alpha) * \mathbf{y}_{1:k-1}^{pre}$ and avoid a computationally intensive non-convex optimization of the full likelihood with respect to $\Delta t_\alpha$ and $\tau_\alpha$.

### Estimating long-term changes in baseline firing rate and synaptic weight

To estimate the time-varying baseline firing rate and the effects of synaptic plasticity ($\beta_{0,k}$, $w_{L,k}$ and $w_{S,k}$) we use two distinct strategies. For the long-term effects we estimate $\beta_{0,k}$ and $w_{L,k}$ by point process adaptive smoothing (Eden et al., 2004; Rauch et al., 1965), and for the short-term effects we estimate $w_{S,k}$ using an additive model that depends on the presynaptic inter-spike intervals (ISIs) (Ghanbari et al., 2017). To estimate all the effects together we use an alternating optimization – we hold the long-term effects constant while updating the short-term effects then hold the short-term effects constant while updating the long-term effects and repeat this alternating pattern until convergence.

#### Estimating Long-Term Effects

To model the long-term changes in baseline firing and synaptic strength we assume that the parameters evolve over time with noisy, linear dynamics. Denoting the parameters as a vector $\boldsymbol{\theta}_k = [\beta_{0,k} \quad w_{L,k}]^T$, we assume that the model parameters evolve over time following

$$\boldsymbol{\theta}_k = \boldsymbol{F}_{k-1}\boldsymbol{\theta}_{k-1} + \boldsymbol{\eta}_k \tag{2.8}$$

where $F_k$ is a system evolution matrix and $\eta_k \sim N(0, Q_k)$ represents Gaussian noise with covariance $Q_k$ at $t_k$. The conditional intensity can be re-written as $\lambda_k = \exp(x_{S,k}^T \theta_k)$ where $x_{S,k} = [1 \quad w_{S,k} \cdot x_k]^T$. We then use adaptive smoothing to track estimates of the parameters given the observed spiking of the pre- and postsynaptic neurons. During a forward step, we first approximate the distribution $p(\theta_k | x_{S,1:k}, y_{1:k})$ using adaptive filtering (Eden et al., 2004). Then, during a backward step, we approximate the distribution $p(\theta_k | x_{S,1:T}, y_{1:T})$ using Rauch-Tung-Striebel (RTS) smoothing (Rauch et al., 1965). In both cases, we approximate the distribution over $\theta_k$ using a multivariate Gaussian.

For adaptive filtering, we assume an initial mean and covariance $\theta_0$ and $W_0$. We first propagate the estimated mean and covariance forward in time according to the process model

$$\theta_{k|k-1} = F_{k-1}\theta_{k-1|k-1} \qquad (2.9)$$
$$W_{k|k-1} = F_{k-1}W_{k-1|k-1}F_{k-1}^T + Q_k \qquad (2.10)$$

Here $\theta_{k|k-1}$ and $W_{k|k-1}$ denote the predicted mean and covariance given observations up to time $t_{k-1}$. We then update the mean and covariance based on the observed spiking at time $t_k$.

$$\lambda_{k|k-1} = \exp(x_{S,k}^T \theta_{k|k-1}) \qquad (2.11)$$
$$W_{k|k}^{-1} = W_{k|k-1}^{-1} + x_{S,k}(\lambda_{k|k-1}\Delta t)x_{S,k}^T \qquad (2.12)$$
$$\theta_{k|k} = \theta_{k|k-1} + W_{k|k}(x_{S,k}(y_k^{post} - \lambda_{k|k-1}\Delta t)) \qquad (2.13)$$

Here $\theta_{k|k}$ and $W_{k|k}$ are the resulting mean and covariance after incorporating the observation at $t_k$. These equations are a special case of updates previously derived for the general Poisson adaptive filtering model (Eden et al., 2004).

Given the estimates from adaptive filtering we then step backwards to find smooth estimates of the parameters. Here we use updates based on the RTS method

$$C_k = W_{k|k}F_k^T W_{k+1|k}^{-1} \qquad (2.14)$$
$$\theta_{k|N} = \theta_{k|k} + C_k(\theta_{k+1|N} - F_k\theta_{k|k}) \qquad (2.15)$$

$$W_{k|N} = W_{k|k} + C_k(W_{k+1|N} - W_{k+1|k})C'_k \qquad (2.16)$$

where $\theta_{k|N}$ and $W_{k|N}$ denote smoothed estimates for the mean and covariance at $t_k$, and to make the algorithm numerically stable, we use an equivalent update

$$\theta_{k|N} = F_k^{-1}(I - Q_{k+1}W_{k+1|k}^{-1})\theta_{k+1|N} + F_k^{-1}(Q_{k+1}W_{k+1|k}^{-1})\theta_{k+1|k} \qquad (2.17)$$

In the results that follow we assume that the process covariance is constant $Q_k = Q$, and that the parameter evolution is a random walk $F_k = I$.

The performance of adaptive smoothing is highly affected by $Q$, and choosing $Q$ improperly can prevent the algorithm from converging. Estimating $Q$ from the data itself using the EM algorithm (Ananthasayanam et al., 2016) is notoriously slow, even with an accelerator (Du & Varadhan, 2020). Here we choose to estimate $Q$ by maximizing prediction likelihood, i.e. likelihood under $\theta_{k|k-1}$. To simplify, we assume that $Q$ is diagonal with independent noise for $\beta_{0,k}$ ($Q_{\beta_0}$) and $w_{L,k}$ ($Q_{w_L}$).

Besides estimating $Q$ by direct two-dimensional optimization, the optimized $Q$ can be approximated by sequential one-dimensional optimization. Since estimation of $Q_{\beta_0}$ depends on $Q_{w_L}$ and vice versa, the order of sequential one-dimensional approximation can influence the results. Since the synaptic effect is sparse ($x_k$) and the time-averaged coupling effect $< x_{S,k} >$ is typically small relative to the baseline, the value of $Q_{w_L}$ has negligible influence on $Q_{\beta_0}$ estimation. Based on this observation, we do one-dimensional approximation by first fixing $Q_{w_L} = 0$ and finding the MLE $\hat{Q}_{\beta_0}$ then fixing $Q_{\beta_0}$ as $\hat{Q}_{\beta_0}$ and finding the MLE $\hat{Q}_{w_L}$.

<u>Estimating Short-Term Effects</u>
In addition to the slow changes in baseline firing and the synaptic weight, we also aim to model fast changes in synaptic weights due to short-term synaptic plasticity. These changes occur on timescales much faster than the typical postsynaptic ISI (~10 ms) and cannot be accurately tracked by adaptive smoothing. Rather than using smoothing, we thus model short-term synaptic plasticity using a parametric model previously introduced in Ghanbari et al. (Ghanbari et al., 2017). Namely, we model the short-term synaptic weight

$$w_{S,k} = 1 + \sum_{\{i|s_i \leq t_k\}} \Delta w_S(\Delta s_i) \exp\left(-\frac{s_i - t_k}{\tau_S}\right) \tag{2.18}$$

$$\Delta w_S(\Delta s_i) = \boldsymbol{b}_S^T(\Delta s_i)\boldsymbol{\alpha}_S \tag{2.19}$$

Here $s_i$ is the $i^{th}$ presynaptic spike time and the $\Delta s_i$ denotes the inter-spike interval (ISI) between the $i^{th}$ and $(i-1)^{th}$ presynaptic spikes. $\Delta w_S(\cdot)$ is a nonlinear function of presynaptic ISI, which describes how the synaptic strength increases or decreases following a pair of presynaptic spikes with a specific ISI. $\Delta w_S < 0$ decreases synaptic strength, mimicking short-term synaptic depression, while $\Delta w_S > 0$ creates increases in synaptic strength akin to facilitation. The cumulative effects of STSP, $w_{S,k}$, are modeled by a convolution of $\Delta w_S(\cdot)$ with the presynaptic spikes, and we assume that the effects decay exponentially with rate $\tau_S$. We parametrize the shape of $\Delta w_S(\cdot)$ using a linear combination of raised-cosine basis functions $\boldsymbol{b}_S^T(\cdot)$. Note that, since we can rewrite the short-term synaptic effect $w_{S,k} = 1 + \tilde{\boldsymbol{x}}_k^T\boldsymbol{\alpha}_S$ where $\tilde{\boldsymbol{x}}_k^T = \sum_{\{i|s_i \leq t_k\}} \boldsymbol{b}_S^T(\Delta s_i) \exp\left(-\frac{s_i - t_k}{\tau_S}\right)$, optimizing $\boldsymbol{\alpha}_S$ (when the long-term effects are fixed) is simply a matter of fitting a GLM.

Additionally, assuming that the long-term effects are fixed allows us to approximate standard errors for the cumulative effects of STSP and the modification function (defined as $1 + \Delta w_S$)

$$Var(w_{S,k}) = \tilde{\boldsymbol{x}}_k^T Var(\boldsymbol{\alpha}_S)\tilde{\boldsymbol{x}}_k \tag{2.20}$$

$$Var(1 + \Delta w_S) = \boldsymbol{b}_S^T Var(\boldsymbol{\alpha}_S)\boldsymbol{b}_S \tag{2.21}$$

To summarize, altogether, we have the following parameters: $\beta_{0,k}$, $\boldsymbol{h}$, $w_{L,k}$, $\boldsymbol{\alpha}_S$, $\Delta t_\alpha$ and $\tau_\alpha$, with hyper-parameters defining the raised-cosine basis $\boldsymbol{b}_S^T(\cdot)$ and timescale $\tau_S$ for short-term modifications, and the long-term weight process covariance $\boldsymbol{Q}$. To fit these parameters, we first estimate the synaptic latency $\Delta t_\alpha$ and time-constant $\tau_\alpha$ directly from the cross-correlogram. Then we optimize the remaining parameters by alternating between fitting the short-term parameters $\boldsymbol{\alpha}_S$ assuming $\beta_{0,k}$, $w_{L,k}$ fixed and fitting the long-term parameters $\beta_{0,k}$, $w_{L,k}$ assuming $\boldsymbol{\alpha}_S$ fixed. Here we assume that $\boldsymbol{h}$ is a fixed, known refractory effect (exponential filter with time constants ranging from 4ms to 10ms), but, in practice, it could easily be estimated alongside the short-term parameters.

Simulations

Here we validate the model using simulated pre- and postsynaptic spiking. If not otherwise specified, presynaptic spike times are generated by a homogeneous Poisson process with a firing rate of 5Hz. The postsynaptic neuron is simulated as a conditionally Poisson process defined in equation 2.3 and 2.4. The observed time length is 20min if not specified, and we use bin size $\Delta t = 1$ms, throughout.

For simulating spike-timing-dependent plasticity (STDP) we use a long-term modification function that depends on the relative timing of pre- and postsynaptic spikes. Here we use a double-exponential modification function, based on the STDP observed in cortical and hippocampal slices (Abbott & Nelson, 2000). In this case, each pair of pre- and post-synaptic spikes modifies the synapse by

$$\Delta w_L(t_k^{pre} - t_k^{post}) = \begin{cases} A_+ \exp\left(\dfrac{t_k^{pre} - t_k^{post}}{\tau_+}\right) & \text{if } t_k^{pre} < t_k^{post} \\[3mm] A_- \exp\left(\dfrac{t_k^{post} - t_k^{pre}}{\tau_-}\right) & \text{if } t_k^{pre} \geq t_k^{post} \end{cases} \qquad (2.22)$$

And we set $A_+ = 0.006$, $A_- = 0.002$, $\tau_+ = 20ms$ and $\tau_- = 20ms$. We further add an additional long-term decay that pushes the synaptic weights back to 1, as in Stevenson and Kording (Stevenson & Kording, 2011). Namely,

$$\begin{aligned} w_L(t_k + \Delta t) = &\ w_L(t_k) - \frac{\Delta t}{\tau_f}(w_L(t_k) - 1) \\ &+ \boldsymbol{I}(y_k^{pre} = 1 \text{ or } y_k^{post} = 1)\Delta w_L(t_k^{pre} - t_k^{post}) \end{aligned} \qquad (2.23)$$

where $\boldsymbol{I}(\cdot)$ is the indicator function, and we set $\tau_f = 20s$.

In the following results, 5 raised-cosine bases with non-linear stretching peaks are used to model STSP modification function within [0, 600ms] in 1ms bins. When fitting the model, we typically assume $Q = I \cdot 10^{-5}$, except in the section "Selection of Hyper-parameter $Q$ in Adaptive Smoothing."

Code for the model and all simulations is available at http://github.com/weigcdsb/GBLM_SMOOTH/.

Experimental Data

Here we analyze two putative synaptic connections from a single large-scale multielectrode array recording shared through the Collaborative Research in Computation Neuroscience Data Sharing Initiative(Ito et al., 2016). Details about the surgery and experiment can be found in (Ito et al., 2014). Briefly, spike trains are spontaneous activity from organotypic slice cultures of mouse somatosensory cortex made using a dense 512-electrode array. Spikes were sorted based on the waveforms of each electrode and it six neighbors using principal component analysis (PCA). Here we detected putative synaptic connections based using the approach in Ren et al. 2020, and fit our model to two strong putative connections. Both connections are drawn from Dataset 23: one from unit 136 to unit 75 (Synapse 1) and one from unit 22 to unit 281 (Synapse 2).

# 3 Results

Here we present simulation results illustrating how this model performs, how the hyperparameters can be efficiently optimized, and why tracking *both* short- and long-term effects can prevent spurious inference of synaptic dynamics. If not otherwise specified, the simulated recording lengths are 20min and presynaptic firing rates are 5Hz.

Simultaneous short-term and long-term changes in synaptic weights

Although most experimental paradigms aim to manipulate and/or observe only a single synaptic timescale, short-term synaptic dynamics (depression and facilitation) and long-term synaptic dynamics (LTD and LTP) coexist and have distinct physiological mechanisms. To illustrate how these dynamics can coexist we simulate an excitatory synaptic connection between a pre- and postsynaptic neuron where the synaptic weight increases instantaneously midway through a 20min recording, while also having short-term synaptic depression on fast timescales (Fig 1). Presynaptic spike timing is generated using a homogeneous Poisson process, while postsynaptic spike times are generated by our full model (Equation 2.3). The overall cross-correlogram between the pre- and postsynaptic shows a short latency, fast onset peak where the probability of postsynaptic spiking increases following each presynaptic spike (Fig 1B, right). Qualitatively, this peak is similar to putative excitatory connections observed in large-scale multielectrode recordings (Barthó et al., 2004), and we can quantify the synaptic strength by estimating the efficacy: the excess probability of a postsynaptic spike occurring following a single presynaptic spike (Levick et al., 1972).

Several previous studies have characterized the short-term and long-term dynamics of spike transmission by partitioning or splitting the overall cross-correlogram to use only a subset of presynaptic spikes. For instance, comparing the cross-correlogram generated using only the presynaptic spikes before the change-point to the cross-correlogram generated using the spikes after the change-point reveals the increase in synaptic strength (Fig 1B). By using presynaptic spikes that are preceded by a specific inter-spike interval we can also observe the effects of short-term synaptic depression (Fig 1D). In this case, presynaptic spikes that occur recently after another spike tend to have lower efficacy compared to presynaptic spikes that occur following a long period of silence, where the synaptic resources have had an opportunity to recover (Fig 1C). Although partitioning the cross-correlogram can reveal clear evidence of long- and short-term changes in spike transmission, it is often unclear what the best partitioning should be and how to combine evidence from multiple types of partitions (e.g. time-based and ISI-based) into a single description. It is also important to note that in the simulation here the firing rate for the presynaptic neuron is set to 5Hz and the firing rate of postsynaptic neuron is ~15Hz. In cases where firing rates are lower and/or when synaptic efficacy is weaker the partitioned cross-correlograms may be too noisy to obtain meaningful estimates of the synaptic weight.

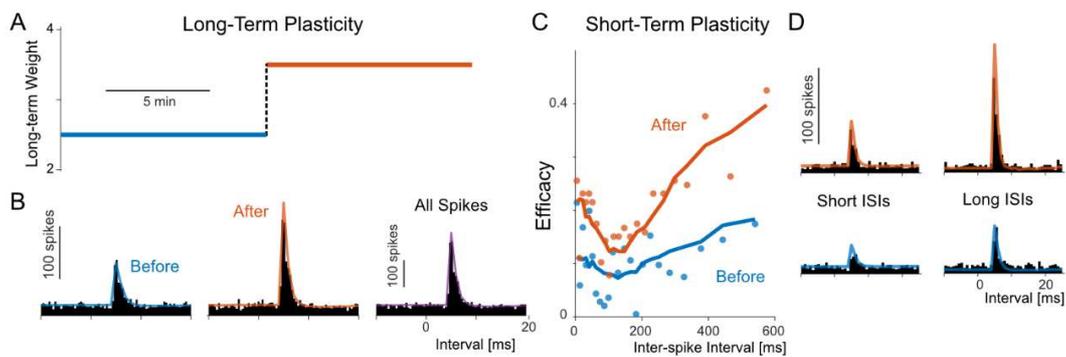

**Fig 1. Characterizing simultaneous short-term and long-term plasticity from spikes.** Here we simulate the spiking of two neurons connected by an excitatory, depressing synapse where the synaptic strength increases abruptly after a change-point **(A)**. We can assess the change in synaptic strength by partitioning the correlogram before (blue) and after (orange) the change-point **(B)**. Here the histograms indicate the observed counts of postsynaptic spiking relative to presynaptic spiking, and the curves denote the corresponding cross-correlation from the model postsynaptic rate. **(C)** By partitioning the correlograms by presynaptic ISI we can also assess the potential influence of short-term synaptic dynamics. Here, shorter presynaptic ISIs correspond to lower postsynaptic efficacies, since the synapse simulated here is depressing. Dots indicate the observed efficacy after splitting the distribution of pre-synaptic ISIs by quantile. Lines indicate the average simulated efficacy from the model itself. **(D)** By partitioning the cross-correlogram based on both the change-point and the presynaptic ISI (median split shown here), we can quantify long-term and short-term synaptic changes simultaneously.

Using a model-based approach the short- and long-term synaptic weights can be estimated simultaneously to create a unified description of postsynaptic spiking. Here we track the long-term changes in synaptic weight using adaptive smoothing, and we fit the short-term changes in synaptic weights using a modification function that describes the additive effects for different presynaptic ISIs (see Methods).

Using adaptive smoothing, the model simply updates its estimate of the synaptic weight based on the observed spike transmission at each time. If postsynaptic spikes follow presynaptic spikes more than expected the parameter for the long-term synaptic weight increases, and if postsynaptic spikes follow presynaptic spikes less than expected the parameter decreases. The hyperparameter $Q_{w_L}$ provides a constraint on how fast the long-term synaptic weight parameter can change. This unstructured approach allows the model to track a wide variety of patterns (Fig 2A). Since the model estimates a synaptic effect for every presynaptic spike, it also provides efficacy estimates for arbitrary partitions of the cross-correlogram (Fig 2B).

For the short-term synaptic weight, we use a more structured approach that depends on the specific pattern of presynaptic ISIs. Following each presynaptic spike, we assume that the short-term synaptic weight increases or decreases by an amount that depends on the inter-spike interval (ISI) preceding that spike. In the absence of presynaptic spikes, we assume that the short-term weight decays (exponentially) back to a baseline weight. These changes can occur rapidly but are constrained by the presynaptic spike timing. Within the model, presynaptic spikes following an interval of 100ms, for instance, will always modify the short-term synaptic weight by the same amount. However, these structured short-term changes allow a range of behaviors, including short-term facilitation and depression-like effects (Fig 2C). These effects occur simultaneously with any long-term changes in synaptic weight, and, again, allow for efficacy estimates with arbitrary partitions of the cross-correlogram (Fig 2D).

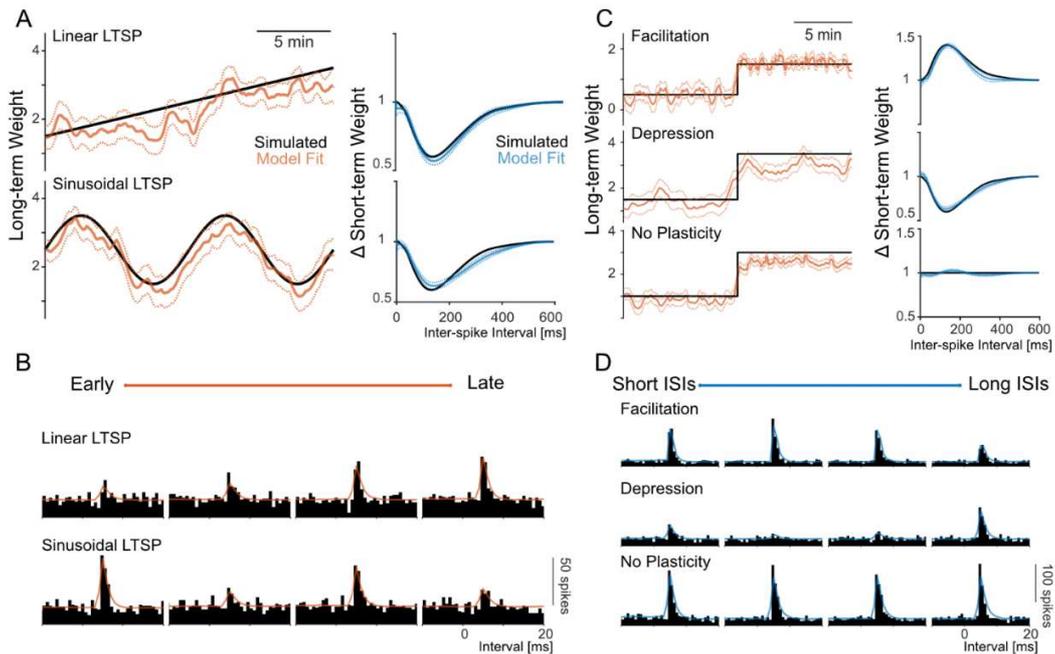

**Figure 2. Tracking long- and short-term plasticity with a model-based approach.** Here we simulate several combinations of short- and long-term changes in synaptic weights. **(A)** shows results from an excitatory, depressing synapse that undergoes long-term changes without a specific change-point. Slow linear (top left, overall efficacy = 0.043) or sinusoidal (bottom left, overall efficacy = 0.049) changes in synaptic weight can be accurately tracked from pre- and postsynaptic spiking alone, and the modification function used to generate short-term dynamics can be accurately reconstructed (right). Here the pre- and postsynaptic baseline firing rates are constant, and dashed lines denote standard error. **(B)** The long-term changes are also apparent when the cross-correlogram is partitioned by recording time. Here the histograms indicate the observed counts of postsynaptic spiking relative to presynaptic spiking, and the curves denote the corresponding cross-correlation from the postsynaptic rates estimated by the model. **(C)** shows results from excitatory synapses that strengthen following a change-point, but with three distinct short-term dynamics: facilitating (top, overall efficacy = 0.12), depressing (middle, overall efficacy = 0.055), or without plasticity (bottom, overall efficacy = 0.19). **(D)** Here the differences in short-term dynamics are apparent when the cross-correlogram is partitioned by presynaptic ISI.

During natural ongoing spiking activity there are unlikely to be large shifts in long-term synaptic strength like Fig 1 and 2. In the absence of external stimulation or task demands, long-term synaptic changes are more likely to result from ongoing patterns of activity. Here we simulate a synapse with spike-timing-dependent plasticity (STDP). The modification function is a traditional double-exponential function, where the synapse is strengthened when postsynaptic spikes follow presynaptic spikes and weakened when presynaptic spikes follow postsynaptic spikes. With the presynaptic spike timing again coming from a homogeneous Poisson process, STDP induces slow fluctuations in the synaptic strength of the full model (Fig 3A). These changes are accurately tracked by the adaptive smoother, even though the smoother does not model pre-post spike timing explicitly. As before, these slow changes also exist simultaneously with short-term synaptic depression (Fig 3B). Although the short-term synaptic effects are visible when partitioning the cross-correlograms (Fig 3C), it is unclear how the long-term fluctuations in STDP could be similarly partitioned.

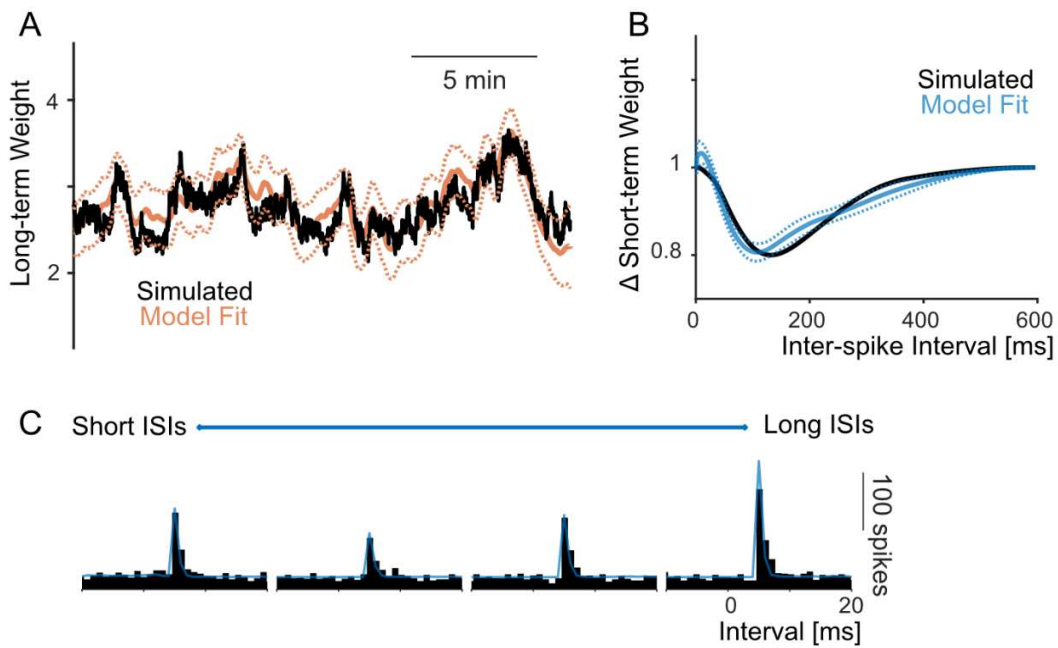

**Figure 3. Tracking naturalistic long-term synaptic dynamics.**
Here we simulate an excitatory, depressing synapse (overall efficacy = 0.13) with long-term changes generated by spike-timing-dependent plasticity (STDP). The model can accurately track the slow fluctuations in synaptic weight induced by a double-exponential STDP learning rule **(A)**, as well as the modification function that generates short-term dynamics **(B)**. As before, the short-term dynamics are also apparent when the cross-correlogram is partitioned by presynaptic ISI. Here the baseline firing rate for postsynaptic neuron is set to ~15Hz, and dashed lines denote standard error.

The previous simulations are all for excitatory synapses. However, the model works similarly for inhibitory synapses. In practice, the "sign" of the synapse is determined by the signs of the long-term and short-term synaptic effects. For simplicity, we assume that the short-term synaptic effect $w_S$ is positive and decays to a baseline value of 1 such that the sign of the long-term effect $w_L$ effectively determines whether the synapse is excitatory $> 0$ or inhibitory $< 0$. As before, both short- and long-term synaptic effects can be estimated when simulating from the full model (Fig 4).

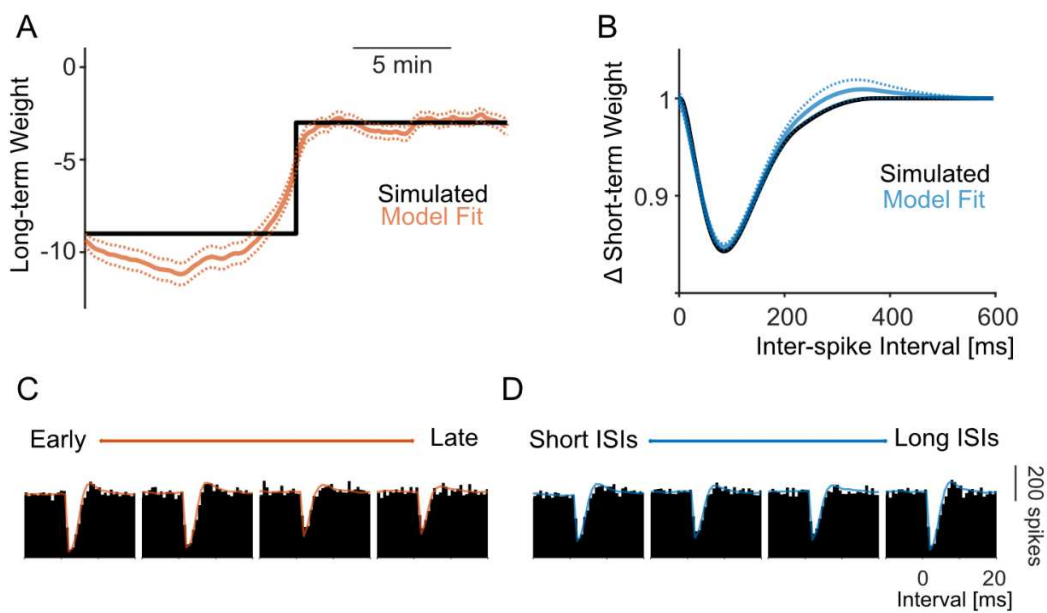

**Figure 4. Tracking the long- and short-term dynamics of an inhibitory synapse.** Here we simulate the spiking of pre- and postsynaptic neurons connected by an inhibitory synapse with depressing short-term dynamics (overall efficacy = -0.06). The firing rate for the presynaptic neuron is 10Hz and the rate for the postsynaptic neuron is ~160Hz. The model accurately tracks the long-term synaptic weight as it weakens following a change-point **(A)**, and accurately fits the modification function that governs short-term depression **(B)**. Dashed lines show standard error. The long- and short-term dynamics are apparent when splitting the cross-correlogram by recording time **(C)** or presynaptic ISIs **(D)**, respectively.

## Variation in pre- and postsynaptic firing rates

In addition to modeling fast and slow changes in synaptic strength we also model slow, unexplained fluctuations in the baseline postsynaptic firing rate. Here we include the baseline parameter $\beta_0$ in the adaptive smoother, and, as with the long-term synaptic weight, it is constrained to vary slowly using the hyperparameter $Q_{\beta_0}$.

To illustrate the effects of a time-varying baseline we simulate, again, a presynaptic neuron with homogeneous Poisson spiking that provides excitatory input to a postsynaptic neuron simulated by the full model. Here the synapse undergoes short-term synaptic depression as well as long-term changes in strength, as before, and, in addition, the firing rate of the postsynaptic neuron fluctuates (Fig 5). In general, the baseline is estimated more accurately and with higher certainty than the long-term synaptic weight. Information about the presence or absence of postsynaptic spikes is always available, but information about the synaptic weight is only available in the short time window following each presynaptic spike. This difference also creates a kind of separability where estimates of the baseline are not particularly influenced by the changes in synaptic weight and estimates of synaptic weight are not influenced by the changing baseline (Fig 5A and C).

As in the previous examples, since there is an estimate of the postsynaptic rate at every time, using a model-based approach allows for reconstruction of arbitrary partitions of the cross-correlogram. When the baseline fluctuates it can also have an influence on the cross-correlogram (Fig 5B and D). Note that here, the baseline affects both the base level of the cross-correlogram as well as the magnitude of the synaptic effect. With the exponential output nonlinearity, the time-varying baseline acts like a time-varying gain and may be useful to mimic slow fluctuations in excitability due to neuromodulation or brain state.

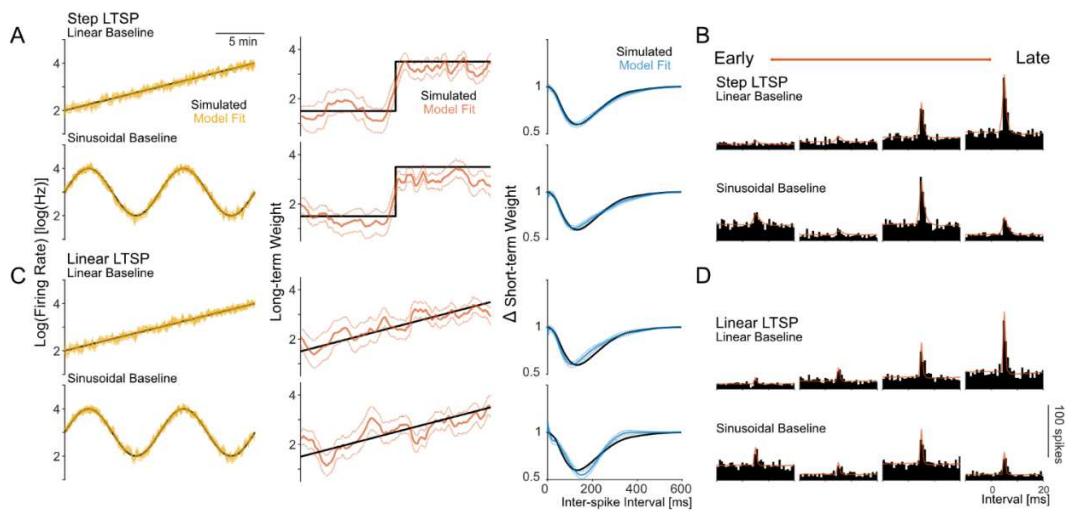

**Figure 5. Accounting for time-varying baseline postsynaptic firing rates.** In addition to tracking variation in synaptic strength, accurately fitting cross-correlograms may also require tracking the postsynaptic rate. Here we simulate an excitatory, depressing synapse that undergoes slow changes in both synaptic strength and postsynaptic rate (overall efficacy = 0.041 to 0.061). The model accurately tracks changes in baseline with sudden **(A)** and continuous **(C)** changes in synaptic strength. Dashed lines denote standard error. In this case, splitting the cross-correlograms by recording time reveals changes in both the baseline correlation and in the amplitude of the peak, and these are well captured by the model (orange lines).

Since information about the synaptic strength is only available when there is a presynaptic spike, the presynaptic firing rate influences how accurately the synaptic weight can be estimated. Higher presynaptic firing rates allow the long-term synaptic strength to be estimated more accurately and with less uncertainty (Fig 6). To illustrate this feature of the model we simulate presynaptic spiking with a Poisson process whose rate changes abruptly, either increasing (Fig 6A) or decreasing (Fig 6B). The postsynaptic neuron spikes according to the full model, in this case, with short-term synaptic facilitation. The baseline postsynaptic rate and long-term synaptic weight are both constant in this simulation. However, the accuracy and precision of the adaptive smoothing estimates for the long-term weight change substantially when the presynaptic rate changes. Estimates for the short-term modification function will also depend on the presynaptic spike rate to some extent, but the impact on accuracy is not as strong, since the short-term model pools information from presynaptic spikes across the entire recording depending on their ISIs.

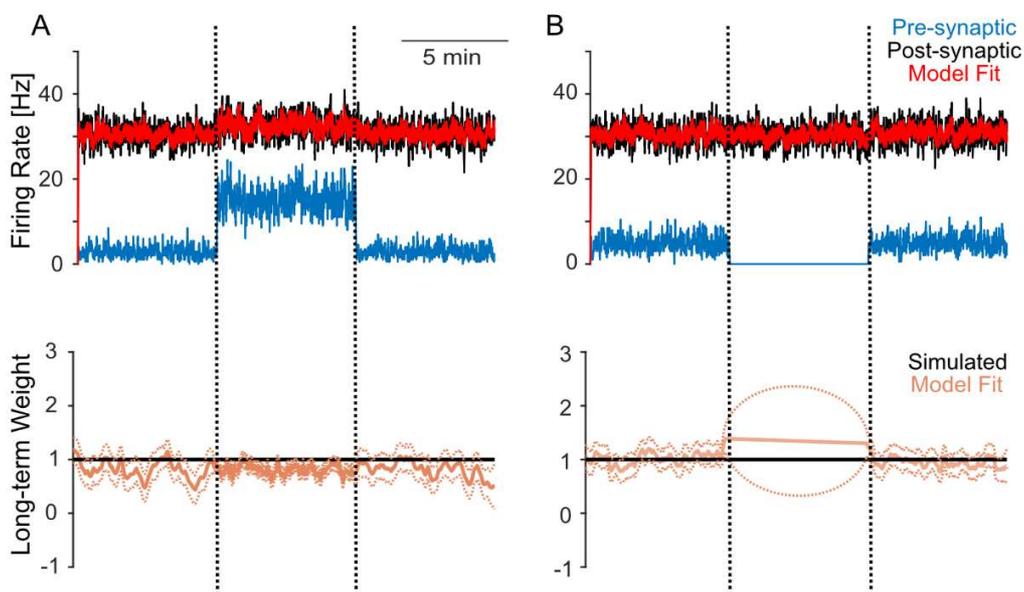

**Figure 6. Presynaptic firing rates influence estimation accuracy.** Here we simulate an excitatory, facilitating synapse with a constant long-term synaptic strength and postsynaptic firing rate (~30Hz). To illustrate how presynaptic firing rates influence the model, the recording time is divided into three time periods with an increase **(A)** or decrease **(B)** in presynaptic rate occurring during the second period. When the presynaptic firing rate increases during the second period (3Hz-15Hz-3Hz), the estimation of the synaptic strength is more precise. In contrast, when the presynaptic firing rate decreases during the second period (5Hz-0Hz-5Hz), the lack of the presynaptic spikes leads to high uncertainty in the synaptic strength. The overall efficacy in (A) is 0.17, and in (B) 0.15. Dashed lines show standard error.

## Omitted variable bias

In the examples above we have shown how estimates of the baseline and synaptic weight are largely separable, since the synaptic effects are constrained to follow presynaptic spikes. Moreover, estimates of the short-term and long-term synaptic effects are also largely separable, since short-term effects are constrained to be ISI-dependent. However, omitting components of the model can result in spurious estimates for the long-term synaptic weight.

To illustrate how omitting effects can bias estimation, we again simulate a presynaptic neuron with Poisson spiking that provides synaptic input to a postsynaptic neuron whose firing is determined by the full model. Here we simulate a synapse with short-term synaptic depression and a constant long-term weight. Although in the previous examples the presynaptic rate has been held constant, in this case we simulate a fluctuating baseline for both the pre- and postsynaptic neurons. Slow fluctuations in the presynaptic rate induce slow changes in the synaptic weight that are purely due to short-term synaptic effects (Fig 7). For the depressing synapse simulated here, a high presynaptic rate causes the synapse to be in a chronically depleted state, while a lower presynaptic rate allows the synapse to recover from depression. In this case, the short-term synaptic weight is negatively correlated with the presynaptic rate.

These slow fluctuations in the short-term synaptic weight are not necessarily problematic and can be accurately estimated with the full model (Fig 7A). However, if we instead fit a model that only includes a long-term synaptic weight and omits the short-term effect, the slow fluctuations are misattributed to the long-term weight (Fig 7B). The adaptive smoother effectively tracks these changes when they are not otherwise explained. Since natural neural activity contains these types of slow fluctuations in presynaptic rate (as a function of brain state, for instance), there are likely to be slow fluctuations in synaptic strength that can be explained away as byproducts of short-term synaptic mechanisms. Modeling short- and long-term effects simultaneously can allow these "explainable" fluctuations to be separated from other phenomena, such as LTP/LTD or STDP.

Omitting the fluctuations in the baseline can also result in spurious estimates of the synaptic strength (Fig 7C). Here, when the model is fit while assuming that the baseline is constant, the adaptive smoother aims to account for higher/lower than expected postsynaptic firing rates using the long-term synaptic weight. Although the synaptic effect is limited to the brief interval following each presynaptic spike, misattributing the rate changes to fluctuations in the synaptic weight improves the overall likelihood. In this case, there is a strong positive correlation between the (omitted) baseline fluctuations and the misestimated long-term synaptic weight. Tracking unexplained variation in the postsynaptic rate is thus likely to be important for accurately tracking synaptic weights in experimental data.

In these scenarios, partitions of the cross-correlogram may serve as a useful check on how well the model describes spike transmission in specific time-periods or as a function of ISI. The model that omits STSP, for instance, will fail to explain the observed depression as a function of ISI, while the model that assumes a constant baseline will fail to model the changing base level of the cross-correlograms when they are partitioned over time.

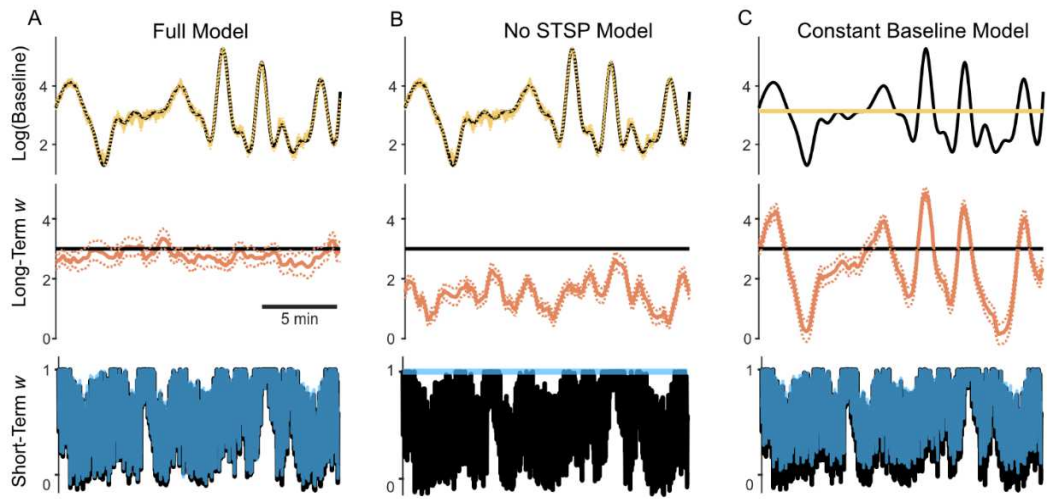

**Figure 7. Omitted variable bias in long-term synaptic weights.** Here we simulate spiking neurons connected by an excitatory, depressing synapse (overall efficacy = 0.13), in this case, with fluctuations in both the baseline postsynaptic firing rate (average 20Hz) and the presynaptic firing rate (average 8Hz). The variation in the presynaptic rate causes slow fluctuations in short-term synaptic strength, but, when fitting the full model **(A)**, these fluctuations are properly attributed to short-term effects. We then consider how omitting variables from the model might influence inference. When all three effects (baseline, LTSP and STSP) are estimated simultaneously they are accurately estimated **(A)**, but when the short-term effects are omitted **(B)** or when the baseline is omitted **(C)** there is substantial misestimation. We find that when the short-term effects are omitted from the model, slow variation in the short-term weight is misattributed to the long-term synaptic effects (B, middle). When the baseline postsynaptic rate is assumed to be constant, the fluctuation in the baseline is also misattributed to long-term synaptic effects (**C**, middle). Dashed lines denote standard error.

Optimizing the adaptive smoother

For the baseline and long-term synaptic weight, the process noise covariance $Q$ defines the timescales for tracking. In the examples above we use a fixed $Q$ during inference. In general, however, $Q$ should be matched to the underlying timescale of the process. If $Q$ is too small the estimated baseline and synaptic strength will be oversmoothed and potentially meaningful changes may be underestimated. On the other hand, if $\boldsymbol{Q}$ is too large the estimated baseline and synaptic strength will be under-smoothed and may reflect noise. Here we assume that $\boldsymbol{Q}$ is diagonal and show how the variance of the baseline $Q_{\beta_0}$ and synaptic weight $Q_{w_L}$ can be optimized by maximizing the prediction likelihood (see Methods).

To illustrate the process of optimizing $\boldsymbol{Q}$ we simulate from the full model and sample the baseline and long-term synaptic weight using a Gaussian random walk with a ground truth $Q_{w_L} = Q_{\beta_0} = 10^{-5}$ . Although the full likelihood always improves with larger values of $Q$, the prediction likelihood $\prod_{k=1}^{N} p(y_k^{post}|\lambda_{k|k-1})$ has a single maximum as a function of $Q_{w_L}$ and $Q_{\beta_0}$ (Fig 8A). We find that changes in $Q_{\beta_0}$ have a much larger effect on the prediction likelihood than changes in $Q_{w_L}$, presumably since the synaptic weight only influences the likelihood in the short interval following each presynaptic spike. Values of $Q_{w_L}$ have negligible influence on prediction likelihood under fixed $Q_{\beta_0}$ but not vice versa. Full 2D optimization of the prediction likelihood (bounded gradient descent in this example) accurately recovers the true process noise, as does a simple 1D optimization scheme where we first optimize $Q_{\beta_0}$ with $Q_{w_L} = 0$ then optimize $Q_{w_L}$ using the $Q_{\beta_0}$ optimized in the first step. As expected, when $\boldsymbol{Q}$ is too high or too low, the estimated baseline and long-term synaptic weights are under- and oversmoothed, respectively (Fig 8B).

Additionally, we find that in a series of short, 10min simulations with pre- and postsynaptic rates set to 5Hz and ~15Hz, respectively, the true values of $\boldsymbol{Q}$ can be accurately recovered (Fig 8C) by maximizing the prediction likelihood. As before, the accuracy in tracking the baseline is typically higher than the accuracy in tracking the synaptic weight. The accuracy in recovering the baseline hyperparameter is also typically higher than the accuracy in recovering the synaptic weight hyperparameter.

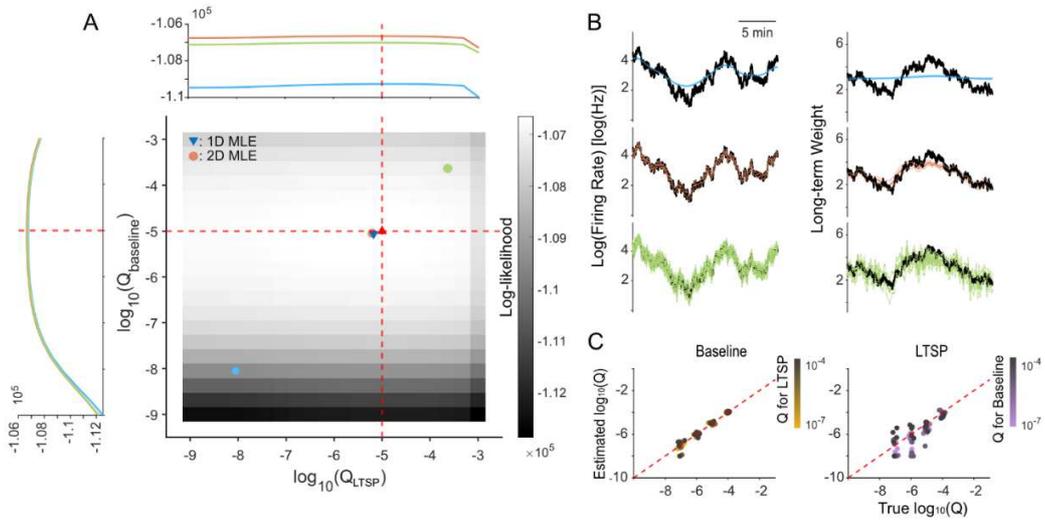

**Figure 8. Selection and influence of $Q$ in point process adaptive smoothing. (A)** Here we simulate an excitatory, depressing synapse with random walk fluctuations in both the long-term synaptic strength and the postsynaptic baseline rate (overall efficacy = 0.17). The heatmap shows the prediction log-likelihood under different values of $Q$. The red triangle and dashed lines denote the true, simulated $Q = 10^{-5}I$. The orange dot represents the maximum prediction likelihood estimate (MLE) $\widehat{Q}_{2D} = Diag(9.0 \times 10^{-6}, 6.2 \times 10^{-6})$ after optimizing the full model. For reference, we also show results when $Q$ is under-estimated (blue, $Q_s = 8.9 \times 10^{-9}I$) and over-estimated (green $Q_L = 2.3 \times 10^{-4}I$). A fast, one-dimensional approximation to the MLE $\widehat{Q}_{1D} = Diag(8.8 \times 10^{-6}, 6.7 \times 10^{-6})$ is also shown (blue triangle). **(B)** shows the corresponding estimates for the baseline and long-term synaptic strength under $Q_s$ (top), $Q_L$ (bottom), and $\widehat{Q}_{2D}$ (middle). Dashed lines denote standard error. When $Q$ is too small, the estimates are over-smoothed; when $Q$ is too large, the estimates are too noisy. We then run many simulations (overall efficacies = 0.14 to 0.30) with different values of $Q$. We find that the MLE (2D) can accurately recover the simulated hyper-parameters over a wide range of values **(C)**. Each combination has 5 replicates and a simulated recording time of 10 min.

## Fitting to Experimental Data

Finally, to illustrate how this model performs with experimental data, we estimated the time-varying weights of two putative synaptic connections. Using data from a one-hour, multi-electrode spike recording of an organotypic culture with mouse somatosensory cortex (see Methods), we find that the model is able to track both short and long-term variation (Fig 9). Here we optimize the process noise covariance $\boldsymbol{Q}$ by maximizing the prediction likelihood for the first 5 minutes of the recording and find $\hat{Q}_{\beta_0} = 1.3 \times 10^{-3}, \hat{Q}_{w_L} = 1.6 \times 10^{-7}$ for Synapse 1 and $\hat{Q}_{\beta_0} = 1.0 \times 10^{-3}, \hat{Q}_{w_L} = 2.5 \times 10^{-6}$ for Synapse 2. After optimizing these hyperparameters we compare the efficacy estimated from the correlograms directly to the model fit. The model is able to track both the efficacy estimated from short, 5 min windows of data (Fig 9A), as well as, the efficacy as a function of the presynaptic ISI (Fig 9G).

The observed efficacy of each of these synapses appears to be a mixture of long- and short-term effects. The model disentangles the overall efficacy into separate long- (Fig 9B) and short-term (Fig 9C) synaptic weights, and also accounts for the substantial variation in baseline postsynaptic firing (Fig 9E). In this case, Synapse 1 appears to be depressing, with higher efficacy at long presynaptic ISIs, while Synapse 2 appears to be facilitating, with higher efficacy at intermediate presynaptic ISIs.

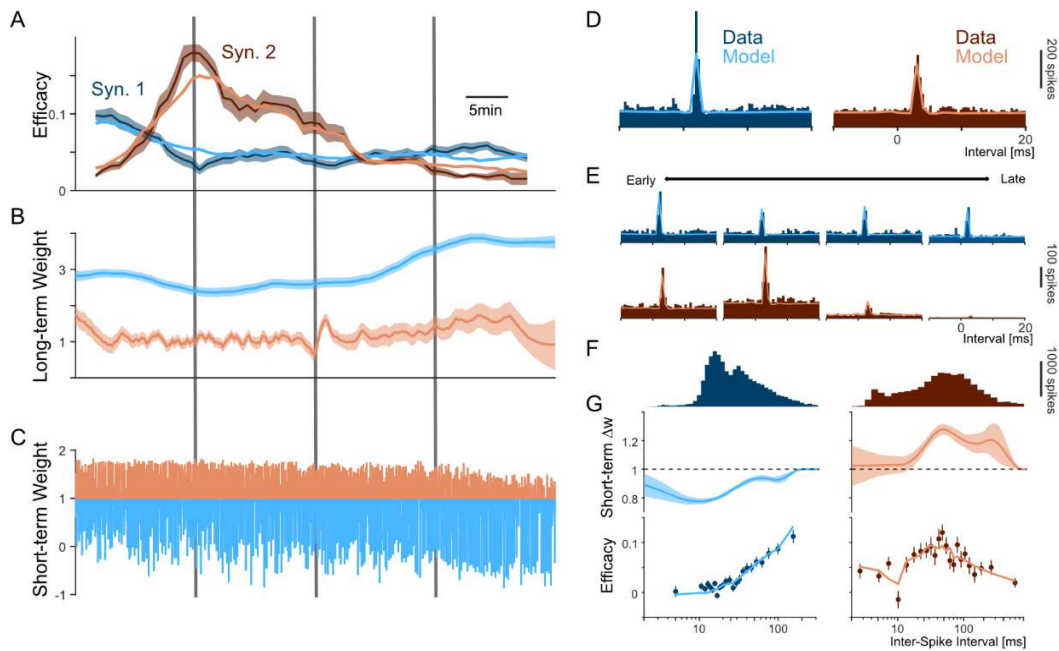

**Fig 9. Application to two putative synaptic connections.** Here we fit our model to pre- and postsynaptic spike trains of two putative synapses from an *in vitro* multielectrode array recording: Synapse 1, with an overall efficacy of 0.037, and Synapse 2, with an overall efficacy of 0.062. **(A)** shows observed and model fitted efficacies estimated in 5 min windows with 4 min overlap. The confidence bands denote the standard error for observed efficacies estimated by block bootstrapping (10 samples). **(B)** and **(C)** show the model long- and short-term weights. By partitioning the overall cross-correlogram **(D)** by recording time, we can see clear changes in the baselines and long-term weights **(E)**. **(G)** shows the short-term modification functions and efficacies according to presynaptic ISI (distribution shown in **F**). Error bars denote standard error of efficacies from i.i.d. bootstrapping of the presynaptic spikes. Confidence bands in **(B)** and **(G)** denote the (conditional) standard error for the model. E

In addition to the model parameters, we also examine the goodness-of-fit for these two synapses. In particular, we compare the log-likelihood ratio of full model (and several reduced models) to a homogeneous Poisson model (Table 1). For both synapses, we find that the full model substantially outperforms a model of a static synapse (with a time-varying baseline but without long- or short-term synaptic weights). Adding LTSP provides only a small improvement relative to the static synapse model, but a model with STSP performs similarly to the full model. This may suggest that the overall variation in these two synapses are mostly due to short-term effects (Fig 9G) and the long-term consequences of these short-term effects (Fig 9C).

|  | Synapse 1 | Synapse 2 |
|---|---|---|
| Full Model | 4.39 | 8.64 |
| Static Synapse | 4.12 | 8.48 |
| LTSP only | 4.13 | 8.49 |
| STSP only | 4.38 | 8.63 |

Table 1. Average log-likelihood ratios for different models (bits/s) relative to a null model (homogenous Poisson). All models have a time-varying baseline postsynaptic rate. The static model has constant LTSP and no STSP. The LTSP only model has time varying LTSP but no STSP. The STSP only model has STSP but no LTSP.

## 4 Discussion

Here we introduce a statistical model that aims to simultaneously track short- and long-term changes in synaptic weights from spike observations. Using simulations, we show that this model can successfully recover time-varying synaptic weights for both excitatory and inhibitory synapses with a variety of short-term and long-term plasticity patterns. We then fit this model to two example putative synapse to illustrate how this model can disentangle the long- and short-term variations in synaptic strength. The accuracy in estimating the long-term synaptic weight depends heavily on the presynaptic firing rate – the more the presynaptic neuron fires, the more accurately synaptic weights can be estimated from spikes. In naturalistic settings, long-term changes in synaptic weights also exist alongside short-term changes in synaptic strength, due to depression or facilitation, and fluctuations in the pre- and postsynaptic rates, due to changes in stimuli, behavior, or brain state. Here we demonstrate why long-term weights, short-term weights, and variations in baseline should all be estimated simultaneously to avoid misinterpretation. If we omit any one of these factors, inference for others will generally be biased.

Including additional covariates can often improve spike prediction accuracy in GLM-like models (Harris et al., 2003; Stevenson, 2018; Truccolo et al., 2005). In these cases, previously unexplained fluctuations in the postsynaptic rate can be explained by an observed covariate such as the firing of simultaneously observed neurons (Harris et al., 2003; Okatan et al., 2005; Pillow et al., 2008), local field potentials (Kelly et al., 2010) or the neuron's own history (Pillow & Simoncelli, 2003; Truccolo et al., 2005). However, once these covariates are considered the baseline firing rate itself is assumed to be constant. Here we consider a general approach where slow fluctuations in the baseline firing rate are directly tracked with adaptive filtering/smoothing. By using adaptive smoothing to track the changing postsynaptic rate, inference of synaptic strengths will be less influenced by unexplained variation caused by slow unobserved variables.

It is important to note, however, that adaptive smoothing cannot track arbitrarily fast changes in the underlying synaptic weight or postsynaptic baseline. In fact, the exact firing rate of a neuron is not necessarily identifiable from single trials (Amarasingham et al., 2015). While adaptive filtering has previously been applied to tracking nonstationary place fields from neurons in the hippocampus (Brown et al., 2001) and for assessing the stability of tuning curves in motor cortex (Stevenson et al., 2011), there may be cases where the assumption of linear dynamics is not appropriate. Additionally, while here we assume a Poisson noise model, spiking activity can be more variable or less variable than Poisson. With stationary models, non-Poisson spiking is typically not a major source of estimation error (Stevenson, 2016). However, the adaptive filtering updates may be sensitive to mis-specified noise, and it may be useful to extend the framework here with other noise models (Gao et al., 2015; Pillow & Scott, 2012).

Additional covariates can also easily be incorporated within the extended-GLM framework used here. Given the increasing scale of simultaneous multi-electrode recordings, it would be particularly interesting to model multiple synaptic inputs simultaneously. Extending the model to allow short- and long-term changes in multiple synaptic inputs would be relatively straightforward. Given $C$ presynaptic neurons, the conditional intensity can be written $\lambda_k = \exp\left(\beta_{0,k} + \boldsymbol{h}\boldsymbol{y}_{k-h:k-1} + \sum_{j=1}^{C} w_k^{(j)} \cdot x_k^{(j)}\right) = \exp\left(\beta_{0,k} + \boldsymbol{h}\boldsymbol{y}_{k-h:k-1} + \sum_{j=1}^{C} w_{L,k}^{(j)} w_{S,k}^{(j)} \cdot x_k^{(j)}\right)$. The long-term weights can be tracked using adaptive smoothing with $1 + C$ parameters: $\boldsymbol{\theta}_k = [\beta_{0,k}, w_{L,k}^{(1)}, \ldots, w_{L,k}^{(C)}]^T$, and the short-term effects can be estimated with a modification function for each synapse. Combining information from a population of spiking neurons, using a state-space model (Paninski et al., 2010; Smith & Brown, 2003), could also potentially allow for fluctuations in the baseline postsynaptic rate to be estimated at timescales faster than what can be estimated from a single neuron alone.

Previous models have aimed to describe short-term changes in synaptic strength from spikes (Chan et al., 2008; English et al., 2017; Ghanbari et al., 2017), to describe long-term changes (Linderman et al., 2014; Robinson et al., 2016; Stevenson & Kording, 2011), or both (Song et al., 2018). Here we use an additive, structured STSP model alongside an unstructured LTSP model, and consider the influence of a changing postsynaptic baseline. While the adaptive smoother simply tracks an estimated synaptic weight based on the observed spiking, it may be preferable to more directly describe the long-term synaptic dynamics with an explicit learning rule based on pre- and postsynaptic spike timing (Song et al., 2018; Stevenson & Kording, 2011) or underlying Calcium dynamics (Graupner & Brunel, 2012). Additionally, while the structured STSP model captures a basic ISI-dependence in the short-term weights, biophysical models of STSP may provide even better descriptions of these dynamics (R. Costa et al., 2013). In some cases, short-term and long-term synaptic plasticity can also interact and tune the sensitive range of synaptic plasticity (R. P. Costa et al., 2015; Deperrois & Graupner, 2020). In previous work, we found that a biophysical model of STSP could account for the short-term dynamics of spike transmission in *in vivo*, experimental recordings (Ghanbari et al., 2020). McKenzie at al. have also found evidence that STSP can generate long-term changes at pyramidal cell-interneuron synapses *in vivo* in the hippocampus (McKenzie et al., 2021). Accounting for long-term changes in the synaptic strength and baseline could further improve models of spike transmission in real data.

The model introduced here aims to track time-varying synaptic weights from simultaneous extracellular recordings from a pre- and postsynaptic neuron. For the sake of simplicity, we assume that monosynaptic connections can be accurately identified, and the simulations here (with efficacies ranging from 0.04 to 0.3) are within the physiological range of efficacies found in a wide range of neural systems (English et al., 2017; Swadlow & Gusev, 2001; Usrey et al., 2000). However, detecting synaptic connections from large-scale multielectrode recordings is not necessarily straightforward, particularly for weak connections or short recording times (Kobayashi et al., 2019; Ren et al., 2020). As experimental methods for verifying the presence of synaptic connections improve, the model proposed here may be useful to extend static models of coupling between neurons to account for potential plasticity. Here we have shown that this model can accurately track variation in two putative synapses. Systematically applying this model to large-scale experimental recordings may lead to new insights into how synapses vary on both short- and long- timescales during ongoing behavior.

## Acknowledgements

This material is based upon work supported by the National Science Foundation under Grant No. 1931249. Thanks to Naixin Ren for helpful comments.

# References

Abbott, L. F., & Nelson, S. B. (2000). Synaptic plasticity: taming the beast. *Nature Neuroscience*, *3*, 1178–1183.

Amarasingham, A., Geman, S., & Harrison, M. T. (2015). Ambiguity and nonidentifiability in the statistical analysis of neural codes. *Proceedings of the National Academy of Sciences*, *112*(20), 6455–6460. https://doi.org/10.1073/PNAS.1506400112

Ananthasayanam, M. R., Mohan, M. S., Naik, N., & Gemson, R. M. O. (2016). A heuristic reference recursive recipe for adaptively tuning the Kalman filter statistics part-1: formulation and simulation studies. *Sādhanā*, *41*(12), 1473–1490. https://doi.org/10.1007/s12046-016-0562-z

Barthó, P., Hirase, H., Monconduit, L., Zugaro, M., Harris, K. D., & Buzsáki, G. (2004). Characterization of neocortical principal cells and interneurons by network interactions and extracellular features. *Journal of Neurophysiology*, *92*(1), 600–608. https://doi.org/10.1152/jn.01170.2003

Brillinger, D. R. (1988). Maximum likelihood analysis of spike trains of interacting nerve cells. *Biological Cybernetics*, *59*(3), 189–200. https://doi.org/10.1007/BF00318010

Brillinger, D. R. (1992). Nerve Cell Spike Train Data Analysis: A Progression of Technique. *Journal of the American Statistical Association*, *87*(418), 260. https://doi.org/10.2307/2290256

Brown, E. N., Nguyen, D. P., Frank, L. M., Wilson, M. A., & Solo, V. (2001). An analysis of neural receptive field plasticity by point process adaptive filtering. *Proceedings of the National Academy of Sciences*, *98*(21), 12261–12266. https://doi.org/10.1073/PNAS.201409398

Carandini, M., Horton, J. C., & Sincich, L. C. (2007). Thalamic filtering of retinal spike trains by postsynaptic summation. *Journal of Vision*, *7*(14), 20–20. https://doi.org/10.1167/7.14.20

Chan, R. H. M., Song, D., & Berger, T. W. (2008). Tracking temporal evolution of nonlinear dynamics in hippocampus using time-varying volterra kernels. *Conference Proceedings : ...*

*Annual International Conference of the IEEE Engineering in Medicine and Biology Society. IEEE Engineering in Medicine and Biology Society. Conference*, *2008*, 4996–4999. https://doi.org/10.1109/IEMBS.2008.4650336

Costa, R. P., Froemke, R. C., Sjöström, P. J., & van Rossum, M. C. W. (2015). Unified pre- and postsynaptic long-term plasticity enables reliable and flexible learning. *ELife*, *4*(AUGUST2015). https://doi.org/10.7554/eLife.09457

Costa, R., Sjöström, P. J., & van Rossum, M. C. W. (2013). Probabilistic inference of short-term synaptic plasticity in neocortical microcircuits. *Frontiers in Computational Neuroscience*, *7*, 75. https://doi.org/10.3389/fncom.2013.00075

Csicsvari, J., Hirase, H., Czurko, A., & Buzsáki, G. (1998). Reliability and state dependence of pyramidal cell-interneuron synapses in the hippocampus: An ensemble approach in the behaving rat. *Neuron*, *21*(1), 179–189. https://doi.org/10.1016/S0896-6273(00)80525-5

Deperrois, N., & Graupner, M. (2020). Short-term depression and long-term plasticity together tune sensitive range of synaptic plasticity. *PLOS Computational Biology*, *16*(9), 1–25. https://doi.org/10.1371/journal.pcbi.1008265

Du, Y., & Varadhan, R. (2020). SQUAREM: An R Package for Off-the-Shelf Acceleration of EM, MM and Other EM-Like Monotone Algorithms. *Journal of Statistical Software; Vol 1, Issue 7 (2020)*. https://doi.org/10.18637/jss.v092.i07

Eden, U. T., Frank, L. M., Barbieri, R., Solo, V., & Brown, E. N. (2004). Dynamic Analysis of Neural Encoding by Point Process Adaptive Filtering. *Neural Computation*, *16*(5), 971–998. https://doi.org/10.1162/089976604773135069

English, D. F., McKenzie, S., Evans, T., Kim, K., Yoon, E., & Buzsáki, G. (2017). Pyramidal Cell-Interneuron Circuit Architecture and Dynamics in Hippocampal Networks. *Neuron*, *96*(2), 505-520.e7. https://doi.org/10.1016/j.neuron.2017.09.033

Fetz, E., Toyama, K., & Smith, W. (1991). *Synaptic Interactions*

*between Cortical Neurons* (pp. 1–47). Springer, Boston, MA. https://doi.org/10.1007/978-1-4615-6622-9_1

Fujisawa, S., Amarasingham, A., Harrison, M. T., & Buzsáki, G. (2008). Behavior-dependent short-term assembly dynamics in the medial prefrontal cortex. *Nature Neuroscience*, *11*(7), 823–833. https://doi.org/10.1038/nn.2134

Gao, Y., Buesing, L., Shenoy, K. V, & Cunningham, J. P. (2015). High-dimensional neural spike train analysis with generalized count linear dynamical systems. In *Advances in Neural Information Processing Systems* (Vol. 28).

Ghanbari, A., Malyshev, A., Volgushev, M., & Stevenson, I. H. (2017). Estimating short-term synaptic plasticity from pre- and postsynaptic spiking. *PLOS Computational Biology*, *13*(9), e1005738. https://doi.org/10.1371/journal.pcbi.1005738

Ghanbari, A., Ren, N., Keine, C., Stoelzel, C., Englitz, B., Swadlow, H. A., & Stevenson, I. H. (2020). Modeling the Short-Term Dynamics of in Vivo Excitatory Spike Transmission. *Journal of Neuroscience*, *40*(21), 4185–4202. https://doi.org/10.1523/JNEUROSCI.1482-19.2020

Graupner, M., & Brunel, N. (2012). Calcium-based plasticity model explains sensitivity of synaptic changes to spike pattern, rate, and dendritic location. *Proceedings of the National Academy of Sciences*, *109*(10), 3991–3996. https://doi.org/10.1073/pnas.1109359109

Harris, K. D., Csicsvari, J., Hirase, H., Dragoi, G., & Buzsáki, G. (2003). Organization of cell assemblies in the hippocampus. *Nature*, *424*(6948), 552–556.

Ito, S., Yeh, F.-C., Timme, N. M., Hottowy, P., Litke, A. M., & Beggs, J. M. (2016). *Spontaneous spiking activity of hundreds of neurons in mouse somatosensory cortex slice cultures recorded using a dense 512 electrode array*. CRCNS.Org. https://doi.org/http://dx.doi.org/10.6080/K07D2S2F

Ito, S., Yeh, F. C., Hiolski, E., Rydygier, P., Gunning, D. E., Hottowy, P., Timme, N., Litke, A. M., & Beggs, J. M. (2014). Large-scale, high-resolution multielectrode-array recording depicts functional network differences of cortical and hippocampal cultures. *PLoS ONE*, *9*(8), 105324. https://doi.org/10.1371/journal.pone.0105324

Kandaswamy, U., Deng, P.-Y., Stevens, C. F., & Klyachko, V. A. (2010). *The Role of Presynaptic Dynamics in Processing of Natural Spike Trains in Hippocampal Synapses*. https://doi.org/10.1523/JNEUROSCI.4050-10.2010

Kelly, R. C., Smith, M. A., Kass, R. E., & Lee, T. S. (2010). Local field potentials indicate network state and account for neuronal response variability. *Journal of Computational Neuroscience*, *29*(3), 567–579. https://doi.org/10.1007/s10827-009-0208-9

Klyachko, V. A., & Stevens, C. F. (2006). Excitatory and Feed-Forward Inhibitory Hippocampal Synapses Work Synergistically as an Adaptive Filter of Natural Spike Trains. *PLoS Biology*, *4*(7), e207. https://doi.org/10.1371/journal.pbio.0040207

Kobayashi, R., Kurita, S., Kurth, A., Kitano, K., Mizuseki, K., Diesmann, M., Richmond, B. J., & Shinomoto, S. (2019). Reconstructing neuronal circuitry from parallel spike trains. *Nature Communications*, *10*(1), 4468. https://doi.org/10.1038/s41467-019-12225-2

Levick, W. R., Cleland, B. G., & Dubin, M. W. (1972). Lateral geniculate neurons of cat: retinal inputs and physiology. *Investigative Ophthalmology*, *11*(5), 302–311. https://pubmed.ncbi.nlm.nih.gov/5028229/

Linderman, S., Stock, C. H., & Adams, R. P. (2014). A framework for studying synaptic plasticity with neural spike train data. In Z. Ghahramani, M. Welling, C. Cortes, N. Lawrence, & K. Q. Weinberger (Eds.), *Advances in Neural Information Processing Systems* (Vol. 27, pp. 2330–2338). Curran Associates, Inc. https://proceedings.neurips.cc/paper/2014/file/4122cb13c7a47 4c1976c9706ae36521d-Paper.pdf

Mantel, G. W. H., & Lemon, R. N. (1987). Cross-correlation reveals facilitation of single motor units in thenar muscles by single corticospinal neurones in the conscious monkey. *Neuroscience Letters*, *77*(1), 113–118. https://doi.org/10.1016/0304-3940(87)90617-3

McKenzie, S., Huszár, R., English, D. F., Kim, K., Christensen, F.,

Yoon, E., & Buzsáki, G. (2021). Preexisting hippocampal network dynamics constrain optogenetically induced place fields. *Neuron*, *109*(6), 1040-1054.e7. https://doi.org/10.1016/j.neuron.2021.01.011

Okatan, M., Wilson, M. A., & Brown, E. N. (2005). Analyzing Functional Connectivity Using a Network Likelihood Model of Ensemble Neural Spiking Activity. *Neural Computation*, *17*(9), 1927–1961.

Paninski, L., Ahmadian, Y., Ferreira, D. G., Koyama, S., Rahnama Rad, K., Vidne, M., Vogelstein, J., & Wu, W. (2010). A new look at state-space models for neural data. In *Journal of Computational Neuroscience* (Vol. 29, Issues 1–2, pp. 107–126). Springer. https://doi.org/10.1007/s10827-009-0179-x

Perkel, D. H., Gerstein, G. L., & Moore, G. P. (1967). Neuronal Spike Trains and Stochastic Point Processes: II. Simultaneous Spike Trains. *Biophysical Journal*, *7*(4), 419–440. https://doi.org/10.1016/S0006-3495(67)86597-4

Pillow, J. W., & Scott, J. G. (2012). Fully Bayesian inference for neural models with negative-binomial spiking. In *Advances in Neural Information Processing Systems* (Vol. 25).

Pillow, J. W., Shlens, J., Paninski, L., Sher, A., Litke, A. M., Chichilnisky, E. J., & Simoncelli, E. P. (2008). Spatio-temporal correlations and visual signalling in a complete neuronal population. *Nature*, *454*(7207), 995–999.

Pillow, J. W., & Simoncelli, E. P. (2003). Biases in white noise analysis due to non-Poisson spike generation. *Neurocomputing*, *52–54*, 109–115. https://doi.org/10.1016/S0925-2312(02)00822-6

Rauch, H. E., Tung, F., & Striebel, C. T. (1965). Maximum likelihood estimates of linear dynamic systems. *AIAA Journal*, *3*(8), 1445–1450. https://doi.org/10.2514/3.3166

Ren, N., Ito, S., Hafizi, H., Beggs, J. M., & Stevenson, I. H. (2020). Model-based detection of putative synaptic connections from spike recordings with latency and type constraints. *Journal of Neurophysiology*. https://doi.org/10.1152/jn.00066.2020

Robinson, B. S., Berger, T. W., & Song, D. (2016). Identification of stable spike-timing-dependent plasticity from spiking activity with generalized multilinear modeling. *Neural Computation*. https://doi.org/10.1162/NECO_a_00883

Smith, A. C., & Brown, E. N. (2003). Estimating a state-space model from point process observations. *Neural Computation*, *15*(5), 965–991. https://doi.org/10.1162/089976603765202622

Song, D., Robinson, B. S., & Berger, T. W. (2018). Identification of Short-Term and Long-Term Functional Synaptic Plasticity From Spiking Activities. In *Adaptive Learning Methods for Nonlinear System Modeling* (pp. 289–312). Elsevier. https://doi.org/10.1016/B978-0-12-812976-0.00017-8

Stevenson, I. H. (2016). Flexible models for spike count data with both over- and under- dispersion. *Journal of Computational Neuroscience*, *41*(1), 29–43. https://doi.org/10.1007/s10827-016-0603-y

Stevenson, I. H. (2018). Omitted Variable Bias in GLMs of Neural Spiking Activity. *Neural Computation*, *30*(12), 3227–3258. https://doi.org/10.1162/neco_a_01138

Stevenson, I. H., Cherian, A., London, B. M., Sachs, N. A., Lindberg, E., Reimer, J., Slutzky, M. W., Hatsopoulos, N. G., Miller, L. E., & Kording, K. P. (2011). Statistical assessment of the stability of neural movement representations. *Journal of Neurophysiology*, *106*(2), 764–774. https://doi.org/10.1152/jn.00626.2010

Stevenson, I. H., & Kording, K. P. (2011). Inferring spike-timing-dependent plasticity from spike train data. In J. Shawe-Taylor, R. S. Zemel, P. Bartlett, F. C. N. Pereira, & K. Q. Weinberger (Eds.), *Advances in Neural Information Processing Systems* (Vol. 24).

Swadlow, H. A., & Gusev, A. G. (2001). The impact of "bursting" thalamic impulses at a neocortical synapse. *Nature Neuroscience*, *4*(4), 402–408. https://doi.org/10.1038/86054

Truccolo, W., Eden, U. T., Fellows, M. R., Donoghue, J. P., & Brown, E. N. (2005). A Point Process Framework for Relating Neural

Spiking Activity to Spiking History, Neural Ensemble, and Extrinsic Covariate Effects. *Journal of Neurophysiology*, *93*(2), 1074–1089.

Usrey, W. M., Alonso, J. M., & Reid, R. C. (2000). Synaptic interactions between thalamic inputs to simple cells in cat visual cortex. *Journal of Neuroscience*, *20*(14), 5461–5467. https://doi.org/10.1523/jneurosci.20-14-05461.2000

Zucker, R. S., & Regehr, W. G. (2002). Short-Term Synaptic Plasticity. *Annual Review of Physiology*, *64*(1), 355–405. https://doi.org/10.1146/annurev.physiol.64.092501.114547